# Using impact exsolution to link the Chicxulub collision and Deccan volcanism

**Kevin G. Harrison**[1]

[1]Geosciences Department, 100 West College Street, PO Box 810, Denison University, Granville, OH 43023-0810 USA; harrisonk@denison.edu; www.KevinGeyerHarrisonPhD.org

January 18, 2017

**Abstract**

Observations suggest that impactors and volcanism are connected, but the mechanism that links these events is unknown. This research proposes the impact exsolution hypothesis to explain how planetary scale collisions trigger volcanism. This hypothesis says that when large objects such as asteroids or comets hit planets, they generate seismic waves that cause exsolution in magma, which initiates eruptions. Observations of the Chicxulub impact and Deccan volcanism support this hypothesis. These observations include changes in feeder dike orientations, changes in isotope geochemistry, and changes in lava flow dynamics following the Chicxulub impact.

**Introduction**

The Chicxulub impact at 66.0 Ma defines the Cretaceous-Paleogene boundary (KPB) and separates two distinct Deccan trap volcanic states (Renne et al., 2015). This study illustrates impact exsolution by showing how the Chicxulub collision triggered a change in Deccan volcanism. The Chicxulub collision generated seismic waves, which initiated exsolution in magma in the mantle. In turn, this exsolution triggered eruptions that produced lava that was not contaminated by the crust. The observed changes in Deccan volcanism following the Chicxulub collision support the impact exsolution hypothesis and help explain the extinction event that occurred at the Cretaceous-Paleogene boundary.

Support for the impact exsolution hypothesis includes the observed differences between pre-impact and post-impact Deccan volcanism (Vanderkluysen et al., 2011; Richards et al., 2015), the discovery of seismic exsolution (Namiki et al., 2016), and estimates of seismic energy generated by the Chicxulub collision (Richards et al., 2015).

**Observed Pre/Post Impact Differences in Deccan Volcanism**

The differences in Deccan volcanism between pre-impact subgroups (Kalsubai and Lonavala) and the post-impact subgroup (Wai), provide experimental evidence for impact exsolution. These differences include dike orientation, lava geochemistry, and eruption dynamics.

The orientation of pre- and post- impact feeder dikes provides insight into the nature of the state shift in Deccan volcanism. The pre-impact Kalsubai subgroup dikes have a North-South orientation, which indicates East-West extensional stresses (Richards et al., 2015). The post-impact Wai subgroup has randomly-oriented dikes (Richards et al., 2015).

The strontium, neodymium, and lead isotopes in the Bushe (pre-impact) formation and the Poladpur (post-impact) formation provide insight into the degree of crustal contamination. The Bushe formation had higher $^{87/86}Sr_t$ values, lower $\varepsilon_{Nd}(t)$ values, and higher $^{206/204}Pb$ values than the Poladpur formation:

Bushe: $^{87/86}Sr_t$ (0.71553 ± 0.00230); $\varepsilon_{Nd}(t)$ (-11.5 ± 5.0); $^{206/204}Pb$ (20.834 ± 1.730) (n=4).

Paladpur: $^{87/86}Sr_t$ (0.70657 ± 0.00131); $\varepsilon_{Nd}(t)$ ( -0.5 ± 3.7); $^{206/204}Pb$ (18.864 ± 0.037) (n=2).

This signifies more crustal contamination in the pre-impact formations than in the post-impact formations. These data were obtained from table 2 in Vanderkluysen et al. (2011). I propose that the state change in Deccan volcanism was caused by seismic exsolution caused by earthquakes generated by the Chicxulub impact.

**Seismic Exsolution**

Namiki et al. (2016) have suggested a general mechanism that links earthquakes and volcanic eruptions: seismic exsolution. Seismic exsolution mobilizes volatiles in magma. Namiki et al. (2016) propose that earthquakes cause exsolution in magma reservoirs, which causes volcanoes to erupt. Building on their seismic exsolution hypothesis, I propose that the Chicxulub impact generated seismic waves that caused the state change in Deccan volcanism described above.

Seismic exsolution occurs when fluid magma oscillates (Namiki et al., 2016). This oscillation occurs in magma reservoirs that are either partially-filled or in magma reservoirs that are full, but have stratified layers. External waves can form in partially-filled magma reservoirs. Internal waves can form in magma that has stratified layers. Seismic waves can cause the magma to exsolve if their frequency is near one of the resonant frequencies of the magma reservoir (Namiki et al., 2016). Low-frequency seismic waves (<1 Hz) are more effective at triggering exsolution than higher-frequency seismic waves (Namiki et al., 2016). Liquid magma is more likely to undergo exsolution if its cross-sectional area is greater than 0.5 meters, but less than 200 meters; if large bubbles are present (e.g., the bubble radius is greater than 1 mm); if the magma has low viscosity (liquid viscosity ($\eta$) < 10 Pascal-second); and if the earthquake is magnitude 9 or greater (Namiki et al., 2016). It follows that magma is more likely to undergo exsolution if the concentration of dissolved volatiles is elevated and if the bubble density is high. Also, escaping bubbles may strip volatiles out as the bubbles rise through the magma, creating a positive feedback loop.

Exsolution may trigger volcanic eruptions in several ways (Namiki, 2016). For example, if:

1) The released gases drive overlying magma to erupt.

2) The exsolved gases transfer heat to the crust, which causes fractures.

3) The build-up of pressure fractures the crust.

4) Exsolution causes the collapse of a foam layer, which causes phenocrysts to fall into deeper layers of magma. These phenocrysts catalyze bubble formation and trigger exsolution in these deeper layers.

5) Depressurization melting may occur, if the exsolved gases escape.

**Seismic Energy Produced by the Chicxulub Impact**

Impact exsolution occurs when large impactors, such as asteroids or comets, generate enough seismic energy to cause seismic exsolution. Impact exsolution requires both serendipity and adequate energy. Serendipity prevails when the seismic waves generated by the impact cause the magma to exsolve if their frequency is near one of the resonant frequencies of magma reservoirs. For magma reservoirs having widths larger than 0.5 meters, energetic waves with long wavelengths will have the greatest effect (Namiki et al., 2016). Large earthquakes, such as the earthquake made by the Chicxulub impact, produced waves like this and had the potential to trigger seismic exsolution. In addition to having a frequency that caused the magma to exsolve, the seismic waves generated by the Chicxulub impact must have had enough energy to cause magma to exsolve. The Chicxulub impact crater has a diameter of 170 km (Arens and West, 2008). Creating such a sizable crater would have generated enough seismic energy to trigger exsolution. For example, earthquakes generated by the Chicxulub impact had estimated magnitudes which ranged from ~9 to ~11 (Richards et al., 2015). In short, it seems possible that the Chicxulub impact provided enough energy at the right wavelength to initiate seismic exsolution.

**Using Impact Exsolution to Explain the Changes between Pre- and Post- Impact Deccan Volcanism**

Changes between pre- and post- impact Deccan volcanism provide support for impact exsolution. The timing of the Chicxulub impact, within 1 to 50,000 years of the state change in Deccan volcanism (Renne et al. 2015), supports the impact exsolution hypothesis. The post-impact eruptions produced over 70% of the Deccan flood basalt volume (Renne et al., 2015). Possible reasons for the observed changes in dike orientation, isotopes, and eruption dynamics in Deccan volcanism following the Chicxulub impact are described below.

*Dike orientation*

As discussed above, the pre-impact dikes have a N-S orientation, which indicates E-W extensional stresses (Richards et al., 2015). The post-impact dikes are randomly-oriented (Richards et al., 2015). Taken together, the dike orientations suggest that regional tectonic extensional stresses played a major role in pre-impact volcanism, but were less important after the impact. Impact exsolution released volatiles from Deccan magma, which may have applied both pressure and heat to the crust. This pressure and heat caused the crust to fracture randomly. The random post-impact dike orientation is consistent with impact exsolution.

*Isotopes*

Post-impact lava had less of a crustal signature than Pre-impact magma (Vanderkluysen et al., 2011). This suggests that seismic waves from the Chicxulub impact triggered exsolution in magma that was located in the mantle. Further, the magma moved from the mantle to the surface with little crustal interaction.

Richards et al. (2015) estimate that the Chicxulub impact generated seismic energy densities of 0.1 - 1.0 $J/m^3$ in the upper ~200 km of mantle. This may have been enough energy to cause exsolution in magma below the Deccan crust.

*Eruption dynamics*

The pre-impact eruption dynamics differed from the post impact eruption dynamics. For example, the pre-impact eruptions were high-frequency and low-volume eruptions (Renne et al., 2015). In contrast, the post-impact eruptions were low-frequency and high-volume (Renne et al., 2015). These low-frequency and high-volume eruptions are consistent with the impact exsolution scenario described below.

**Impact Exsolution Deccan Scenario**

Richards et al. (2015) have suggested that the Chicxulub impact caused a state change in Deccan volcanism. Renne et al. (2015) have concluded that this state change in volcanism occurred within 50,000 years of the Chicxulub collision, which provides further support for the link between the impact and the change in volcanism. However, the mechanism that links the two events was unclear until Namiki et al. (2016) discovered seismic exsolution. In short, the seismic waves generated by the Chicxulub impact triggered exsolution in magma located below the crust, which caused the observed state shift in Deccan volcanism. One possible scenario:

1) Volatiles escaped from the magma located below the crust due to seismic exsolution.

2) The escaped gases migrated upward and delivered heat and pressure to the crust, which caused it to fracture randomly.

3) The resulting depressurization, caused by the release of volatiles, enabled additional magma to melt, caused more exsolution, and created areas of low pressure above the magma.

4) Liquid magma invaded the low-pressure areas and erupted.

5) Magma moved in from below to replace the erupted magma, was exposed to lower pressures, and released volatiles. This would re-start the cycle at step 1.

**Discussion**

If earthquakes can trigger volcanic eruptions by generating seismic waves that cause exsolution (Namiki et al., 2016) and impactors cause earthquakes, then impactors may trigger volcanic eruptions. If impactors trigger volcanic eruptions, there are at least two implications:

- By generating seismic waves that cause exsolution in magma chambers, impactors may trigger volcanic eruptions that range from continental flood basalts to super volcanoes.

- Impactors may have triggered past volcanic eruptions on Earth and other planets, and may play a role in triggering future volcanic eruptions.

Furthermore, the co-occurrence of impactor collisions and volcanism have been linked with mass extinction (Arens and West, 2008; Peterson et al., 2016). The combined effects of the impact and flood volcanism associated with the Chicxulub collision and Deccan volcanism have created one of the largest mass extinction events in the geological record (Arens and West, 2008). Also, the frequency of impactors may have been underestimated (Speyerer et al., 2016).

**Conclusion**

Impact exsolution may help explain the state shift in Deccan volcanism following the Chicxulub impact. Both theory and experimental observations back this hypothesis. Theoretical support includes the discovery of seismic exsolution and estimates of seismic activity following the Chicxulub impact. The Chicxulub impact generated seismic waves, which caused volatiles in magma to exsolve. These exsolved volatiles fractured the crust and escaped to the atmosphere. The resulting decompression extended below the crust and melted magma, which erupted directly to the surface. This scenario can account for the observed changes in Deccan volcanism following the Chicxulub impact. The impact exsolution hypothesis may help improve our understanding of the events surrounding the KPB by linking the Chicxulub impact with the state shift in Deccan volcanism. Impacts may have triggered past volcanic eruptions on Earth and other planets, and may play a role in triggering future volcanic eruptions.

**Acknowledgments**

KGH thanks BAZ and EGS.